\documentstyle[psfig]{elsart}
\begin{document}
\begin{frontmatter}
\title{Simple Models of the Protein Folding Problem}
\author{Chao Tang\thanksref{talk}}
\address{NEC Research Institute, 4 Independence Way, Princeton, New
Jersey 08540, USA}
\thanks[talk]{Invited talk at Dynamics Days Asian Pacific, Hong Kong,
July 13-16, 1999. To appear in Physica A.}

\begin{abstract}
The protein folding problem has attracted an increasing attention 
from physicists. The problem has a flavor of statistical mechanics, but
possesses the most common feature of most biological problems -- the
profound effects of evolution. I will give an introduction to the problem,
and then focus on some recent work concerning the so-called ``designability
principle''.
The designability of a structure is measured by the number of sequences
that have that structure as their unique ground state. Structures differ
drastically in terms of their designability; highly designable structures
emerge with a number of associated sequences much larger than the average.
These highly designable structures 1) possess ``proteinlike'' secondary
structures and motifs, 2) are thermodynamically more stable, and 3) fold
faster than other structures. These results suggest that protein
structures are selected in nature because they are readily designed and
stable against mutations, and that such selection simultaneously leads to
thermodynamic stability and foldability. According to this picture, a key
to the protein folding problem is to understand the emergence and the
properties of the highly designable structures.

\vspace{0.4cm}
\vspace{-0.5cm}
\end{abstract}
\end{frontmatter}

\section{Introduction}
\begin{figure}
\centerline{\psfig{file=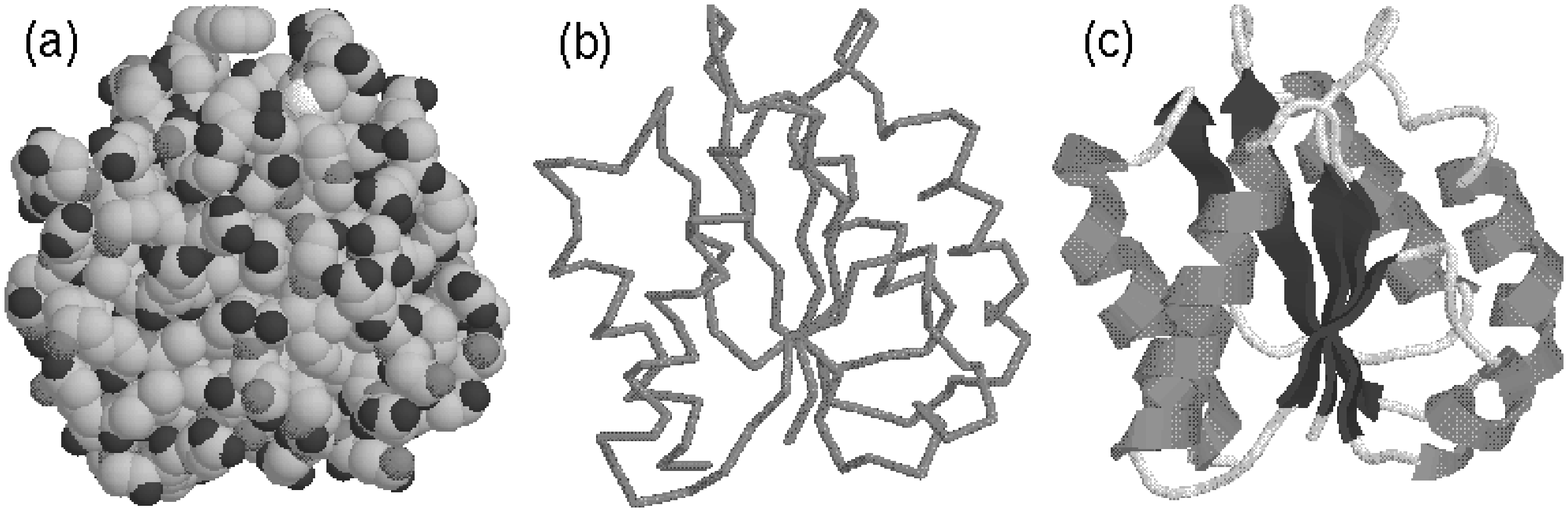,width=5.5in}}
\vspace{0.2cm}
\caption{The protein {\it flavodoxin}. (a) The atomic model. Different
atoms are shown in different grey scales: oxygen (dark),
nitrogen (dark grey), carbon (light grey), and sulfur (white).
Hydrogen atoms are not shown.
(b) The backbone of the structure which is the formed by connecting the
$C_\alpha$ atoms of each amino acid along the chain.
(c) The ribbon diagram of the backbone. $\beta$-strand is shown in dark,
$\alpha$-helix in grey, and turns and loops in white.}
\label{flavo}
\end{figure}

The word ``protein'' originates from the Greek word {\it proteios} which
means ``of the first rank'' \cite{protein}. Indeed, proteins are building
blocks and functional units of all biological systems. They play crucial
roles in virtually all biological processes. Their diverse functions
include enzymatic catalysis, transport and storage, coordinated motion,
mechanical support, signal transduction, control and regulation, and
immune response. A protein consists of a chain of amino acids whose
sequence is determined by the information in DNA/RNA. There are 20
natural amino acids nature uses to make up proteins. These differ in size
and other physical and chemical properties. The most important difference
however, as far as the determination of the structure is concerned, is
their hydrophobicity, i.e. how much they dislike water. An open protein
chain, under normal physiological conditions, will fold into a
three-dimensional configuration to perform its function. This folded
functional state of the protein is called the {\it native state}. For
single domain globular proteins which
are our focus here, the length of the chain is of the order of 100 amino
acids (from $\sim 30$ to $\sim 400$). Proteins of longer chains usually
form multidomains each of which can usually fold independently. In
Fig.~\ref{flavo} is shown a globular protein {\it flavodoxin} whose function
is to transport electrons. Like most water soluble single domain globular
proteins, it is very compact with a roughly rounded shape. The folded
geometry of the chain can be best viewed in the cartoonish ribbon diagram
(Fig.~\ref{flavo}c) of the backbone configuration (Fig.~\ref{flavo}b).
One can see that the geometry of this protein structure is several 
$\alpha$-helices sandwiching an $\beta$-sheet. The folded geometries, often
referred to as {\it folds}, of proteins usually look far more regular than
random, typically possessing secondary structures (e.g., $\alpha$-helices
and $\beta$-sheets) and sometime even having tertiary symmetries. (One can
recognize an approximate mirror symmetry in Fig.~\ref{flavo}c.) One of
the main goals of the protein folding problem is to predict the
three-dimensional folded structure for a given sequence of amino acids. 

\begin{figure}
\centerline{\psfig{file=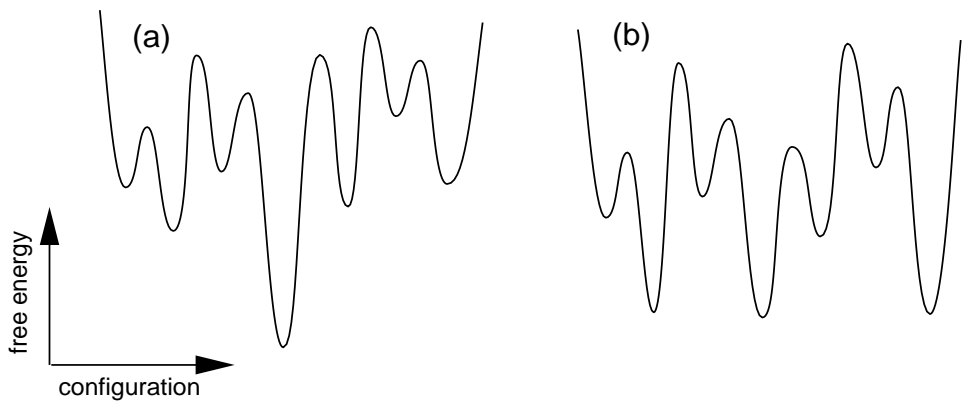,width=5in}}
\caption{The schematic energy landscapes of (a) a protein sequence and
(b) a random sequence.}
\label{lands}
\end{figure}

The protein folding problem is the kind of biological problem that has
an immediate appeal to physicists. A protein can fold (to its native
state) and unfold (to a flexible open chain) reversibly by changing the
temperature, pH, or the concentration of some denaturant in solution.
While the study of denaturation of proteins can be traced back at least
70 years when Wu \cite{wu} pointed out that denaturation was in fact the
unfolding of the protein from ``the regular arrangement of a rigid
structure to the irregular, diffuse arrangement of the flexible open
chain''. A turning point was the work of Anfinsen on the so-called 
``thermodynamic hypothesis'' in the late 50s and early 60s. Anfinsen
\cite{anfinsen} and later many others demonstrated that for single
domain proteins 1) the information coded in the amino acid sequence of
a protein completely determines its folded structure, and 2) the native
state is the global minimum of the free energy. These conclusions should
be somewhat surprising to physicists. For the configurational ``(free)
energy landscape'' of a heteropolymer of the size of a protein is typically
``rough'', in the sense that there are typically many metastable states some
of which have energies very close to the global minimum. How could a protein
always fold into its unique native state with the lowest energy? The answer
is evolution. Indeed, random sequences of amino acids are usually ``glassy''
and usually can not fold uniquely. But natural proteins are not random
sequences. They are a small family of sequences, selected by nature via
evolution, that have a distinct global minimum well separated from other
metastable states (Fig.~\ref{lands}). One might ask: what are the unique
and yet common properties of this special ensemble of proteinlike
sequences? In other words, can one distinguish them from other sequences
without the arguably impossible task of constructing the entire energy
landscape? The answer lies in the heart of the question we introduce in the
next paragraph and is the focus of this discussion.

\begin{figure}
\centerline{\psfig{file=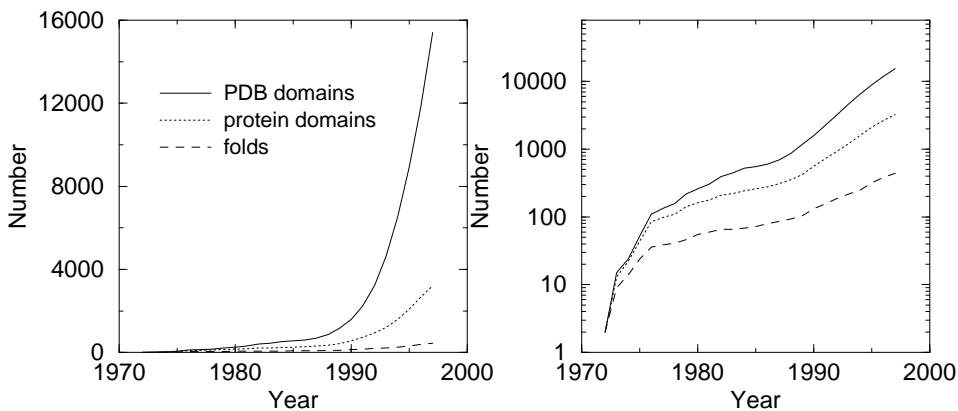,width=5in}}
\caption{The cumulative numbers of PDB domains, (non-redundant) protein
domains, and folds vs. year. Source: SCOP [4] and Ref. [6]. Courtesy of
Dr. Steven Brenner.}
\label{scop}
\end{figure}

\begin{figure}
\centerline{\psfig{file=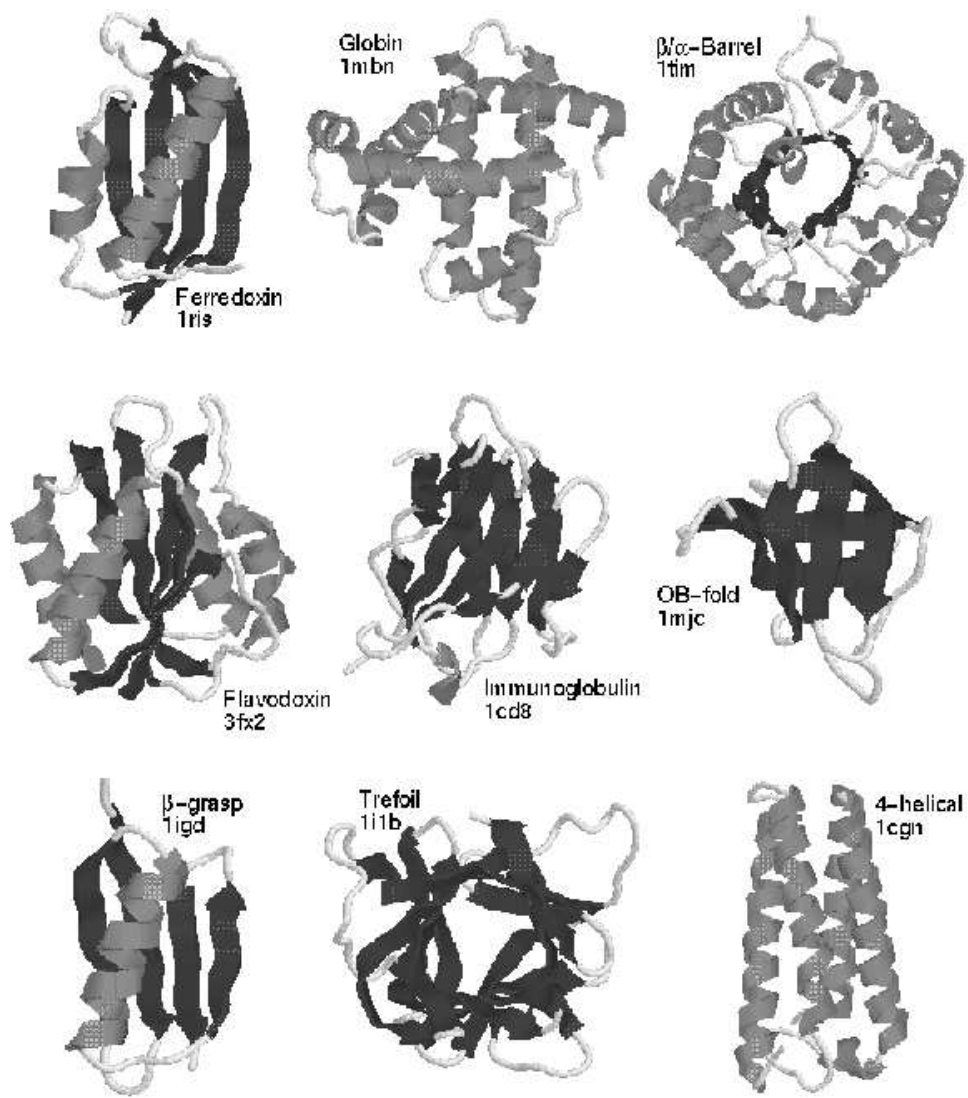,width=5in}}
\caption{Representatives of some popular folds.}
\label{folds}
\end{figure}

There are about 100,000 different proteins in the human body. The number is
much larger if we consider all natural proteins in the biological world.
Protein structures are classified into different {\it folds}. Proteins of
the same fold have the same major secondary structures in the same
arrangement with the same topological connections \cite{scopsite}, with
some small variations typically in the loop region. So in some sense,
folds are distinct templates of protein structures. Proteins with a
close evolutionary relation often have high sequence similarity and share
a common fold. What is intriguing is that common folds occur even for
proteins with different evolutionary origins and biological functions. The
number of folds is therefore much lower than the number of proteins. 
Shown in Fig.~\ref{scop} is the cumulative number of solved protein domains
\cite{footnote} along with the cumulative number of folds as a function of
the year. It is increasingly less likely that a newly solved protein
structure would take a new fold. It is estimated that the total number of
folds for all natural proteins is only about 1000 \cite{brenner,cho-fold}.
Some of the frequently observed folds, or ``superfolds'' \cite{orengo}, are
shown in Fig.~\ref{folds}. Among apparent features of these folds are
secondary structures, regularities, and symmetries. Therefore, as in the
case of sequences, protein structures or folds are
also a very special class. One might ask: Is there anything special about
natural protein folds--are they merely an arbitrary outcome of evolution or
is there some fundamental reason behind their selection \cite{foot-des}? Is
the selection of protein structures coupled with the selection of protein
sequences? We will now address these questions via a thorough study of
simple models. 

\section{Simple Models and the Designability}
\begin{figure}
\centerline{\psfig{file=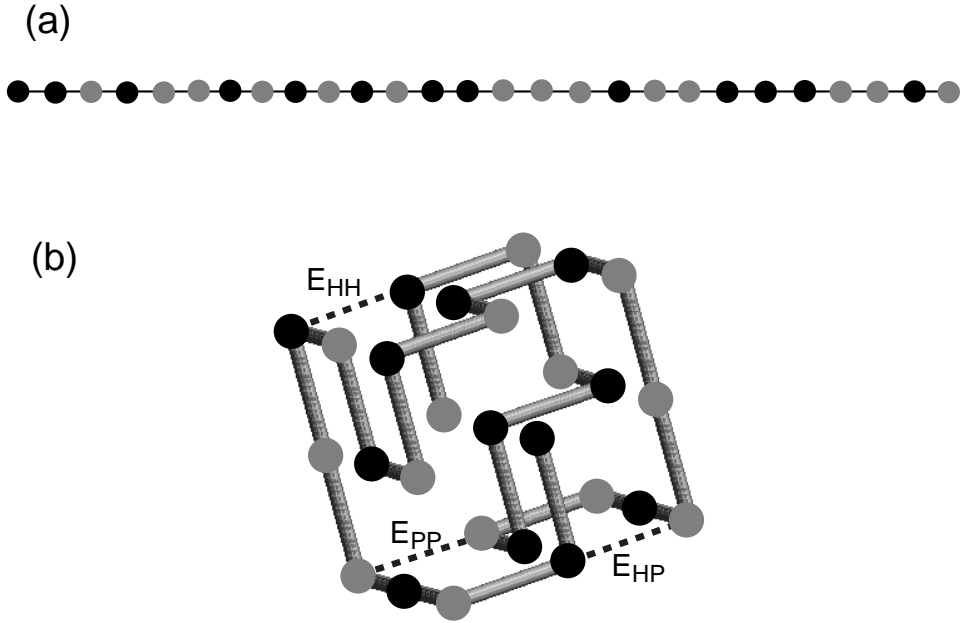,width=5in}}
\caption{A 3D lattice HP model. A sequence of H (dark disc) and P (light
disc) (a)
is folded into a 3D structure (b).}
\label{3d}
\end{figure}

The dominant driving force for protein folding is the so-called hydrophobic
force \cite{hydro}. The 20 amino acids differ in their hydrophobicity and
can be very roughly classified into two groups: hydrophobic and polar
\cite{ltw}. Hydrophobic amino acids have greasy side chains made of
hydrocarbons and like to stick together in water to minimize their contact
with water. Polar amino acids have polar groups (with oxygen or nitrigen) in
their side chains and do not mind so much to be in contact with water. The
simplest model of protein folding is the so-called ``HP lattice model''
\cite{hp}, whose structures are defined on a lattice and whose sequences
take only two ``amino acids'': H (hydrophobic) and P (polar) (see
Fig.~\ref{3d}). The energy for a sequence folded into a structure is
simply given by the short range contact interactions
\begin{equation}
H=\sum_{i<j}e_{\nu_i\nu_j}\Delta ({\bf r}_i-{\bf r}_j),
\label{ham}
\end{equation}
where $\Delta ({\bf r}_i-{\bf r}_j)=1$ if ${\bf r}_i$ and ${\bf r}_j$ 
are adjoining lattice sites but $i$ and $j$ are not adjacent in position 
along the sequence, and $\Delta ({\bf r}_i-{\bf r}_j)=0$ otherwise.
Depending on the types of monomers in contact, the interaction energy
$e_{\nu_i\nu_j}$ will be $e_{\rm HH}$, $e_{\rm HP}$, or $e_{\rm PP}$,
corresponding to H-H, H-P, or P-P contacts, respectively (see 
Fig.~\ref{3d}) \cite{foot-e}. We \cite{lhtw} choose these interaction
parameters to
satisfy the following physical constraints: 1) compact shapes have lower
energies than any non-compact shapes; 2) H monomers are buried as much as
possible, expressed by the relation $e_{\rm PP}>e_{\rm HP}>e_{\rm HH}$, 
which lowers the energy of configurations in which Hs are hidden from 
water; 3) different types of monomers tends to segregate, expressed by 
$2 e_{\rm HP}>e_{\rm PP}+e_{\rm HH}$. Conditions 2) and 3) were derived 
from the analysis \cite{ltw} of the real protein data contained in the 
Miyazawa-Jernigan matrix \cite{mj} of inter-residue contact energies
between different types of amino acids. Since we consider only the
compact structures all of which have the same total number of contacts,
we can freely shift and rescale the interaction energies, leaving only one
free parameter. Throughout this section, we choose $e_{\rm HH}=-2.3$,
$e_{\rm HP}=-1$ and $e_{\rm PP}=0$ which satisfy conditions 2) and 3) above.
The results are insensitive to the value of $e_{\rm HH}$ as long
as both these conditions are satisfied. (The analysis in Ref.~\cite{ltw}
on the interaction potential of amino acids arrived at a form 
\begin{equation}
e_{\mu\nu}=h_{\mu}+h_{\nu}+c(\mu,\nu),
\label{emj}
\end{equation}
where $h_{\mu}$ is the hydrophobicity of the amino acid $\mu$ and $c$ is
a small mixing term. The additive term, i.e. the hydrophobic force,
dominates the potential. The choice of $e_{\rm HH}=-2.3$ in our study can
be viewed as a result of a hydrophobic part -2 plus a small mixing part
-0.3. Several authors have investigated the effect of the mixing
contribution as a small perturbation to the additive potential
\cite{guo,reza}.)

\begin{figure}
\centerline{\psfig{file=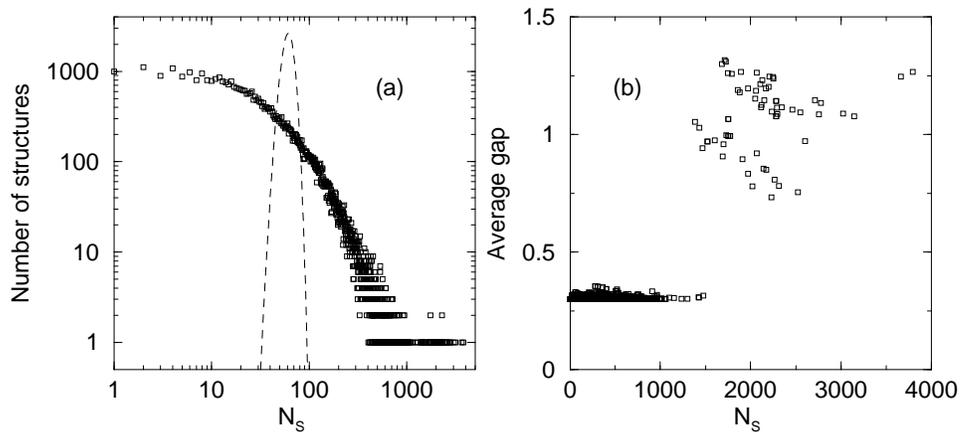,width=5in}}
\caption{(a) Histogram of $N_S$ for the $3\times3\times3$ system. (b)
Average energy gap between the ground state and the first excited
state vs. $N_S$ for the $3\times3\times3$ system.}
\label{histo3d}
\end{figure}

We have studied the model (\ref{ham}) on a three dimensional cubic lattice
and on a two dimensional square lattice \cite{lhtw}. 
For the three dimensional case, we analyze a chain composed of 27 monomers. 
We consider all the structures which form a compact $3\times 3\times 3$
cube. There are a total of 51,704 such structures unrelated by rotational,
reflection, or reverse labeling symmetries \cite{stru333,lhtw}. 
For a given sequence, the ground state structure is found by calculating
the energies of all compact structures. We completely enumerate the ground
states of all $2^{27}$ possible sequences. We find that only $4.75\%$ of the
sequences have unique ground states and thus are potential proteinlike
sequences. We then calculate the designability of each compact structure.
Specifically, we count the number of sequences $N_S$ that have a given
compact structure $S$ as their unique ground state. We find that compact
structures differ drastically in terms of their designability, $N_S$.
There are structures that can be designed by an enormous number of
sequences, and there are ``poor'' structures which can only be designed by
a few or even no sequences. For example, the top structure can be designed
by $3,794$ different sequences ($N_S=3,794$), while there are $4,256$
structures for which $N_S=0$. The number of structures
having a given $N_S$ decreases monotonically (with small fluctuations)
as $N_S$ increases (Fig.~\ref{histo3d}a). There is a long tail to the 
distribution. Structures contributing to the tail of the distribution
have $N_S>>\overline {N_S}=61.7$, where $\overline {N_S}$ is the average
number.  We call these structures ``highly designable'' structures. The
distribution is very different from the Poisson distribution (also shown
in Fig.~\ref{histo3d}a) that would result if the compact structures were
statistically equivalent. For a Poisson distribution with a mean
$\overline {N_S}=61.7$, the probability of finding even one structure
with $N_S>120$ is $1.76\times 10^{-6}$.

The highly designable structures are, on average, thermodynamically more
stable than other structures. The stability of a structure can be
characterized by the average energy gap $\overline {\delta_S}$, averaged
over the $N_S$ sequences that design the structure. For a given sequence,
the energy gap $\delta_S$ is defined as the minimum energy difference
between the ground state energy and the energy of a different compact
structure. We find that there is a marked correlation between $N_S$ and
$\overline{\delta_S}$ (Fig.~\ref{histo3d}b). Highly designable structures
have average gaps much larger than those of structures with small $N_S$,
and there is a sudden jump in $\overline{\delta_S}$ for structures with
$N_S^c \approx 1,400$. This jump is a result of two different kinds of
excitations a ground state could have. One is to break an H-H bond and a
P-P bond to form two H-P bonds, which has an (mixing) energy cost of
$2E_{\rm HP}-E_{\rm HH}-E_{\rm PP}=0.3$. The other is to change the
position of an H-mer from relatively buried to relatively exposed, so the
number of H-water bonds (the lattice sites outside the $3\times3\times3$
cube are occupied by water molecules) is increased. This kind of
excitations has an energy $\ge 1$. The jump in Fig.~\ref{histo3d}b
indicates that the lowest excitations are of the first kind for $N_S<N_S^c$,
but are a mixture of the first and the second kind for $N_S>N_S^c$.

\begin{figure}
\centerline{\psfig{file=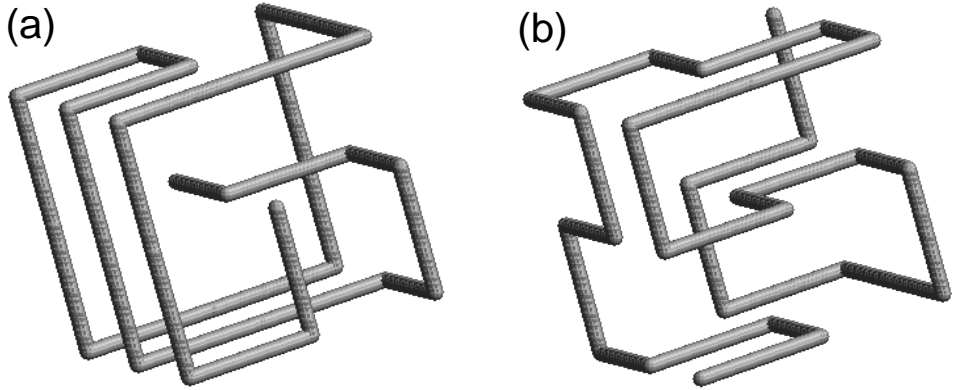,width=5in}}
\caption{The top structure (a) and an ordinary structure with $N_S=1$ (b)
for the $3\times3\times3$ system.}
\label{top3d}
\end{figure}

A striking feature of the highly designable structures is that they exhibit
certain geometrical regularities that are absent from random structures and
are reminiscent of the secondary structures in natural proteins. In
Fig.~\ref{top3d} is shown the most designable structure along with a
typical random structure. We examined the compact structures with the 10
largest $N_S$ values and found that all have parallel running lines folded
in a regular manner. 

\begin{figure}
\centerline{\psfig{file=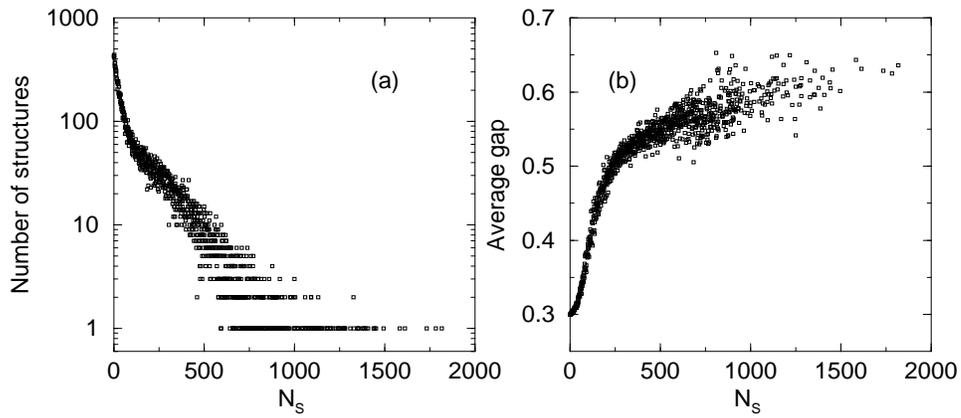,width=5in}}
\caption{Histogram of $N_S$ (a), and the average energy gap between
the ground state and the first excited state vs. $N_S$ (b), for the 2D
$6\times6$ HP model.}
\label{histo2d}
\end{figure}

\begin{figure}
\centerline{\psfig{file=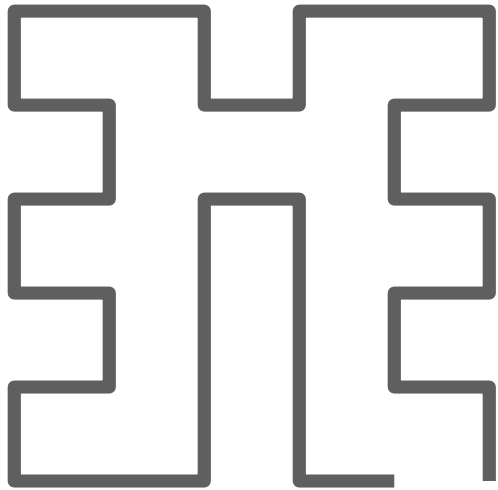,width=2.5in}}
\caption{The top structure for the 2D $6\times6$ system.}
\label{top2d}
\end{figure}

We have also studied the model on a 2D lattice. We take sequences of length
36 and fold them into compact $6\times6$ structures on the square lattice.
There are $28,728$ such structures unrelated by symmetries including the
reverse-labeling symmetry. In this case, we did not enumerate all $2^{36}$
sequences but randomly sampled them to the
extend where the histogram for $N_S$'s reached a reliable distribution.
Similar to the 3D case, the $N_S$'s have a very broad distribution
(Fig.~\ref{histo2d}a). In this case the tail decays more like an
exponential. The average gap also correlates positively with $N_S$
(Fig.~\ref{histo2d}b). Again
similar to the 3D case, we observe that the highly designable structures
in 2D also exhibit secondary structures. In the 2D $6\times 6$ case, as
the surface-to-interior ratio approaches that of real proteins, the highly
designable structures often have bundles of pleats and long strands,
reminiscent of $\alpha$ helices and $\beta$ strands in real proteins; in
addition, some of the highly designable structures have tertiary
symmetries (Fig.~\ref{top2d}).

\begin{figure}
\centerline{\psfig{file=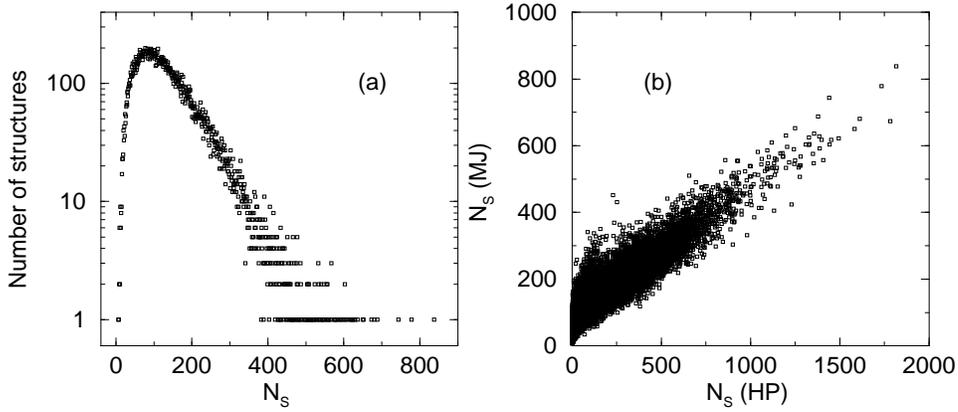,width=5in}}
\caption{(a) Histogram of $N_S$ for the 2D $6\times6$ model with the MJ
matrix, obtained with $3,990,000$ random sequences. (b) $N_S$ from the HP
model vs. $N_S$ from the MJ matrix for 2D $6\times6$ structures.}
\label{20lett}
\end{figure}

To ensure that the above results are not an artifact of the HP model, we
have studied model (\ref{ham}) with 20 amino acids \cite{ltw-20}. In this
case the interaction energies $e_{\nu_i\nu_j}$, where $\nu_i$ can now be
any one of the 20 amino acids, are taken from the Miyazawa-Jernigan matrix
\cite{mj}--an empirical potential between amino acids. For the 3D
$3\times3\times3$ system and the 2D $6\times6$ system, the total numbers of
sequences are $20^{27}$ and $20^{36}$, respectively, which are impossible to
enumerate. So we randomly sampled the sequence space. Similar to the case
of the HP model, $N_S$'s have a broad distribution in both 3D and 2D cases.
Furthermore, the $N_S$'s correlate well with the ones obtained from the HP
model (see Fig.~\ref{20lett}). Thus the highly designable structures in the
HP model are also highly designable in the 20-letter model
\cite{foot-rich}. With 20 amino
acids, there are few sequences that will have {\it exactly} degenerate
ground states. For example, in the case of $3\times3\times3$ about $96.7\%$
of the sequences have unique ground states. However, many of these ground
states are almost degenerate, in the sense that there are compact structures
other than the ground state with energies very close to the ground state
energy. If we require that for a ground state to be truly unique there
should be no other states of energies within $g_c$ from the ground state
energy, then the percentage of the sequences that have unique ground
states is reduced to about $30\%$ and $8\%$ for $g_c=0.4k_B$T and
$g_c=0.8k_B$T, respectively. 

\section{A Geometrical Interpretation}
A number of questions arise: Among the large number of structures, why are
some structures highly designable? Why does designability also guarantee
thermodynamic stability? Why do highly designable structures have
geometrical regularities and even symmetries? In this section we address
these questions by using a geometrical formulation of the protein folding
problem \cite{ltw-pnas}.

As we have mentioned before, the dominant driving force for protein folding
is the hydrophobicity, i.e. the tendency for hydrophobic amino acids to hide
away from water. To model only the hydrophobic force in protein folding, 
one can assign parameters $h_\nu$ to characterize the hydrophobicities of
each of the 20 amino acids \cite{hydroph}. Each sequence of amino acids then
has an associated vector 
${\bf
h}=(h_{\nu_1},h_{\nu_2},\ldots,h_{\nu_i},\ldots,h_{\nu_N})$,
where $\nu_i$ specifies the amino acid at position $i$ of the sequence. 
The energy of a sequence folded into a particular structure is taken to be
the sum of the contributions from each amino acid upon burial away from
water: 
\begin{equation}
H=-\sum_{i=1}^N s_i h_{\nu_i},
\label{ham1}
\end{equation}
where $s_i$ is a structure-dependent number characterizing the degree of
burial of the $i$-th amino acid in the chain. Eq.~(\ref{ham1}) is
essentially a solvation model \cite{eis} at the residue level and can also
be obtained by taking the mixing term of Eq.~(\ref{emj}) to zero
\cite{guo,reza,ltw-pnas}.

\begin{figure}
\centerline{\psfig{file=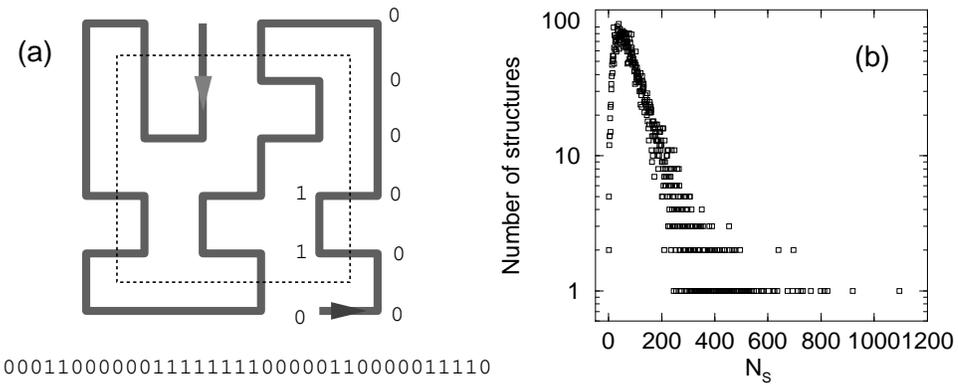,width=5in}}
\caption{(a) A $6\times6$ compact structure and its corresponding string.
A structure is represented by a string $\vec{s}$ of 0s and 1s,
according to whether a site is on the surface or in the core (which is
enclosed by the dotted lines), respectively. (In fact, two structures,
related by the reverse-labeling symmetry, are shown, corresponding to
the two opposite paths indicated by the two arrows. So the $\vec{s}$ of
one structure is the reverse of the other.)
(b) The histogram of $N_S$ for the $6\times6$ PH model obtained by
using 19,492,200 randomly chosen sequences. 
}
\label{site-type}
\end{figure}

To simplify the discussion, let us consider only compact structures and
let $s_i$ take only two values: 0 and 1, depending on whether the amino acid
is on the surface or in the core of the structure, respectively. Therefore,
each compact structure can be represented by a string $\{s_i\}$ of 0s and
1s: $s_i=0$ if the $i$-th amino acid is on the surface and $s_i=1$ if it is
in the core (see Fig.~\ref{site-type}a for an example on a lattice). Let us
make further simplification by using only two amino acids: $\nu_i={\rm H}$ or
P, and let $h_{\rm H}=1$ and $h_{\rm P}=0$. Thus a sequence $\{\nu_i\}$ is
also mapped into a string $\{\sigma_i\}$ of 0s and 1s: $\sigma_i=1$ if
$\nu_i={\rm H}$ and $\sigma_i=0$ if $\nu_i={\rm P}$. Let us call this model
the PH (Purely Hydrophobic) model.
Assuming every compact structure of a given size has the same numbers of
surface and core sites and noting that the term $\sum_i {\sigma_i}^2$ is a
constant for a fixed sequence of amino acids and does not play any role in
determining the relative energies of structures folded by the sequence, 
Eq.~(\ref{ham1}) is then equivalent to \cite{ltw-pnas}:
\begin{equation}
H = \sum_{i=1}^N (\sigma_i - s_i)^2.
\label{ham2}
\end{equation}
Therefore, the energy for a sequence $\vec{\sigma}=\{\sigma_i\}$ folded onto
a structure $\vec{s}=\{s_i\}$ is simply the distance squared (or the Hamming
distance in the case where both $\{\sigma_i\}$ and $\{s_i\}$ are strings
of 0s and 1s) between the two vectors $\vec{\sigma}$ and $\vec{s}$.

We can now formulate the designability question geometrically.
We have two ensembles or spaces: one being all the
sequences $\{\vec{\sigma}\}$ and the other all the structures $\{\vec{s}\}$.
Both are represented by $N$-dimensional points or vectors where $N$ is the
length of the chain. The points of all the sequences are trivially
distributed in the $N$-dimensional space. In the case of the PH model, the
points representing sequences are all the vertices of an $N$-dimensional
hypercube (all possible $2^N$ strings of 0s and 1s of length $N$). On the
other hand, the points representing all the structures $\{\vec{s}\}$ have a
very different distribution in the $N$-dimensional space. The $\vec{s}$'s
are constrainted and correlated. For example, in the case of the PH model
where $s_i=0$ or 1, not every string of 0s and 1s actually represents a
structure. In fact, only a very small fraction of the $2^N$ strings of 0s
and 1s correspond to structures. If we consider only compact structures where
$\sum_i s_i=n_c$ with $n_c$ the number of core sites, then the structure
vectors $\{\vec{s}\}$ cover only a small fraction of the vertices of a
hyperplane in the $N$-dimensional hypercube.

\begin{figure}
\centerline{\psfig{file=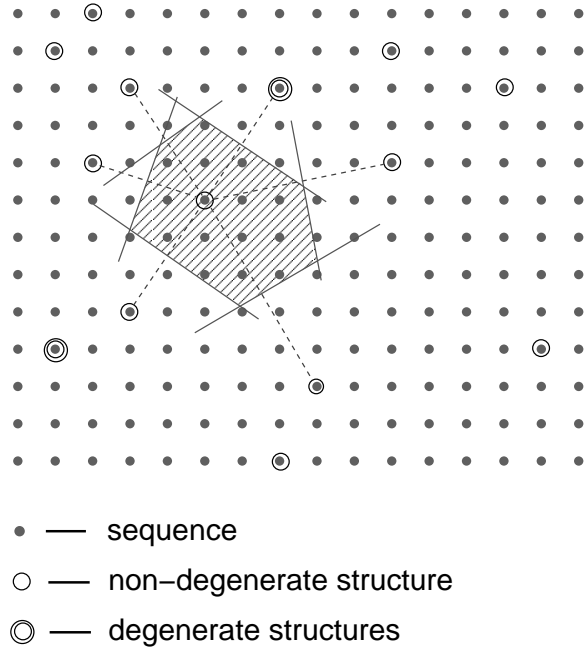,width=3in}}
\caption{The sequence and structure ensembles in $N$-dimension.}
\label{voron}
\end{figure}

Now imagine putting all the sequences $\{\vec{\sigma}\}$ and all the
structures $\{\vec{s}\}$ together in the $N$-dimensional space (see
Fig.~\ref{voron} for a schematic illustration). (In a more general case
it would be simplest to picture if one normalizes $\{\vec{h}\}$ and
$\{\vec{s}\}$ so that $0\le h_i,s_i \le 1$.) From Eq.~(\ref{ham2}), it is
evident that a sequence will have a structure as its unique ground state if
and only if the sequence is closer (measured by the distance defined by
Eq.~(\ref{ham2})) to the structure than to any other structures. Therefore,
the set of all sequences $\{\vec{\sigma}(\vec{s})\}$ that uniquely design a
structure $\vec{s}$ can be found by the following geometrical construction:
Draw bisector planes between $\vec{s}$ and all of its neighboring structures
in the $N$-dimensional space (see Fig.~\ref{voron}). The volume enclosed by
these planes is called the Voronoi polytope around $\vec{s}$. 
$\{\vec{\sigma}(\vec{s})\}$ then consists of all sequences within the
Voronoi polytope. Hence, the designabilities of structures are directly
related to the distribution of the structures in the $N$-dimensional space.
A structure closely surrounded by many neighbors will have a small Voronoi
polytope and hence a low designability; while a structure far away from
others will have a large Voronoi polytope and hence a high designability.
Furthermore, the thermodynamic stability of a folded structure is directly
related to the size of its Voronoi polytope. For a sequence $\vec{\sigma}$,
the energy gap between the ground state and an excited state is the
difference of the squared distances between $\vec{\sigma}$ and the two
states (Eq.~(\ref{ham2})). A larger Voronoi polytope implies, on average, a
larger gap as excited states can only lie outside of the Voronoi polytope
of the ground state. Thus, this geometrical representation of the problem
naturally explains the positive correlation between the thermodynamic
stability and the designability.

\begin{figure}
\centerline{\psfig{file=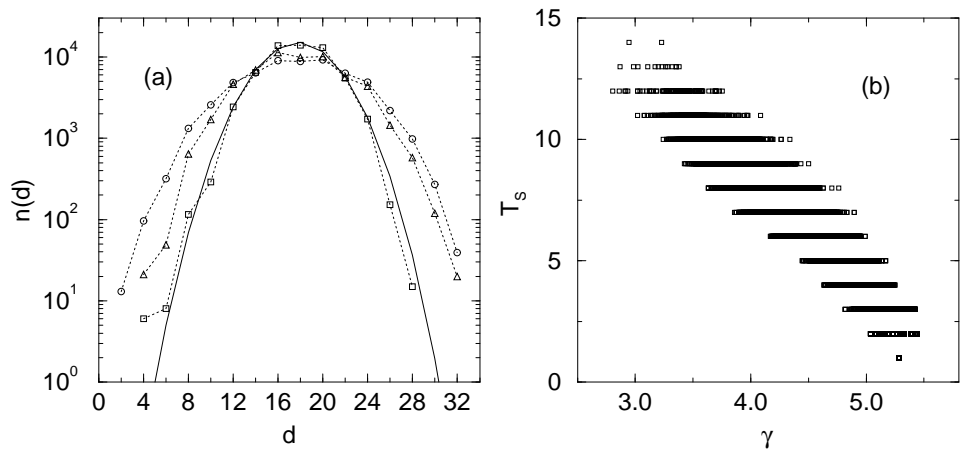,width=5in}}
\caption{(a) Number of structures vs. the Hamming distance
for three structures with low (circles), intermediate (triangles) and high
(squares) designability. Also plotted is $n^0(d)$ (solid line).
(b) The number of transitions between core and surface sites
vs. $\gamma$ for all the $6\times6$ compact structures.}
\label{nd}
\end{figure}

As a concrete example, we have studied a 2D PH model of $6\times6$
\cite{ltw-pnas}. For each compact structure, we devide the 36 sites into 16
core sites and 20 surface sites (see Fig.~\ref{site-type}a). Among the
$28,728$ compact structures unrelated by symmetries, there are $119$ that
are reverse-labeling symmetric. (For a reverse-labeling symmetric structure,
$s_i=s_{N+1-i}$.) So the total number of structures a sequence can fold
onto is $(28,728 - 119)\times 2 + 119 = 57,337$, which map into $30,408$
distinct strings. There are cases in which two or more structures map into
the same string. We call these structures degenerate structures and a
degenerate structure can not be the unique ground state for any sequence
in the PH model. Out of the $28,728$ structures, there are $9,141$
nondegenerate structures (or $18,213$ out of $57,337$). A histogram for
the designability of
nondegenerate structures is obtained by sampling the sequence space using
19,492,200 randomly chosen sequences and is shown in Fig.~\ref{site-type}b.
The set of highly designable structures are essentially the same as those
obtained from the HP model discussed in the previous section. To further
probe how structure vectors are distributed in the $N$-dimensional space, we
measure the number of structures, $n_{\vec{s}}(d)$, at a Hamming distance
$d$ from a given structure $\vec{s}$. Note that all the $57,337$ structures
are distributed on the vertices of the hyperplane
defined by $\sum_i s_i=16$. There are a total of $C_{36}^{16}=7,307,872,110$
vertices in the hyperplane. If the structure vectors were distributed
uniformly on these vertices, $n_{\vec{s}}(d)$ would be the same for all
structures and would be: $n^0(d)=\rho N(d)$, where
$\rho=57,337/7,307,872,110$ is the average density of structures on
the hyperplane and $N(d)=C_{16}^{d/2}C_{20}^{d/2}$ is the number of vertices
at distance $d$ from a given vertex. In Fig.~\ref{nd}a,
$n_{\vec{s}}(d)$ is plotted for three different structures with low,
intermediate, and high designabilities, respectively, along with $n^0(d)$.
We see that a highly designable structure typically has fewer neighbors than
a less designable structure, not only at the smallest $d$s but out to $d$s
of order 10-12. Also, $n_{\vec{s}}(d)$ is considerably larger than $n^0(d)$
for small $d$ for structures with low designability. These results
indicate that the structures are very nonuniformly distributed and are
clustered--there are highly populated regions and lowly populated regions. 
A quantitative measure of the clustering environment around a structure is
the second moment of $n_{\vec{s}}(d)$,
\begin{equation}
\gamma^2(\vec{s})=<d^2>-<d>^2=4\sum_{ij}s_is_jc_{ij},
\label{gamma}
\end{equation}
where
\begin{equation}
c_{ij}=<s_is_j>-<s_i><s_j>
\label{cij}
\end{equation}
and $<\cdot>$ denotes average over all structures. In Fig.~\ref{zscore}a
we plot the designability $N_S$ of a structure vs. its $\gamma$. We see
that while larger $N_S$
implies smaller $\gamma$ it is not true vice versa. This is because that
$N_S$ is very sensitive to the local environment at small $d$s while $\gamma$
is more a global measure.

What are the geometrical characteristics of the structures in the highly
populated regions and lowly populated regions, respectively? This is
something we are very interested in but know very little about. Naively,
the structures in the highly populated regions are typical random structures
which can be easily transformed from one to another by small local changes.
On the other hand, structures in lowly populated regions are ``atypical''
structures which tend to be more regular and ``rigid''. They have fewer
neighbors so it is harder to transform them to other structures with only
small rearrangements. One geometrical feature of highly designable structures
is that they have more surface-to-core transitions along the backbone, i.e.
there are more transitions between 0s and 1s in the structure string for a
highly designable structure than average \cite{ltw-pnas}. We found a good
correlation between the number of surface-core transitions in a structure
string $\vec{s}$, $T(\vec{s})$, and $\gamma(\vec{s})$ (Fig.~\ref{nd}b).
Thus, a necessary condition for a structure to be highly designable is to
have a small $\gamma$ or a large $T$.

A great advantage of the PH model is that it is simple enough to test some
ideas immediately. Two quantities often used to characterize structures are
the energy spectra ${\mathcal{N}}(E,\vec{s})$ \cite{fink,shak99} and
${\mathcal{N}}(E,\vec{s},C)$ \cite{shak99}. The first one is the energy
spectrum of a given structure, $\vec{s}$ over all sequences,
$\{\vec{\sigma}\}$:
\begin{equation}
{\mathcal{N}}(E,\vec{s})=\sum_{\{\vec{\sigma}\}}
\delta[H(\vec{\sigma},\vec{s})-E].
\label{spec}
\end{equation} 
The second one is over all sequences of a fixed composition $C$ (e.g. fixed
numbers of H-mers and P-mers in the case of two-letter code), 
$\{\vec{\sigma}\}_C$:
\begin{equation}
{\mathcal{N}}(E,\vec{s},C)=\sum_{\{\vec{\sigma}\}_C}
\delta[H(\vec{\sigma},\vec{s})-E].
\label{spec-c}
\end{equation}
It is easy to see that if two structure strings $\{s_i\}$ and
$\{s^\prime_i\}$ are related by permutation, i.e. $s_i=s^\prime_{k_i}$,
for $i=1,2,\cdots,N$, where $k_1,k_2,\cdots,k_N$ is a permutation of
$1,2,\cdots,N$, then ${\mathcal{N}}(E,\vec{s})={\mathcal{N}}(E,\vec{s^\prime})$
and ${\mathcal{N}}(E,\vec{s},C)={\mathcal{N}}(E,\vec{s^\prime},C)$. Thus
all maximally compact structures have the same energy spectra
Eqs.~(\ref{spec}) and (\ref{spec-c}). Therefore, structures differ in
designability, not because they have different energy spectra Eqs.~(\ref{spec})
and (\ref{spec-c}) \cite{fink,shak99}, but because they have different
neighborhood in the structure space.

\section{Folding Dynamics and Thermodynamic Stability}
\begin{figure}
\centerline{\psfig{file=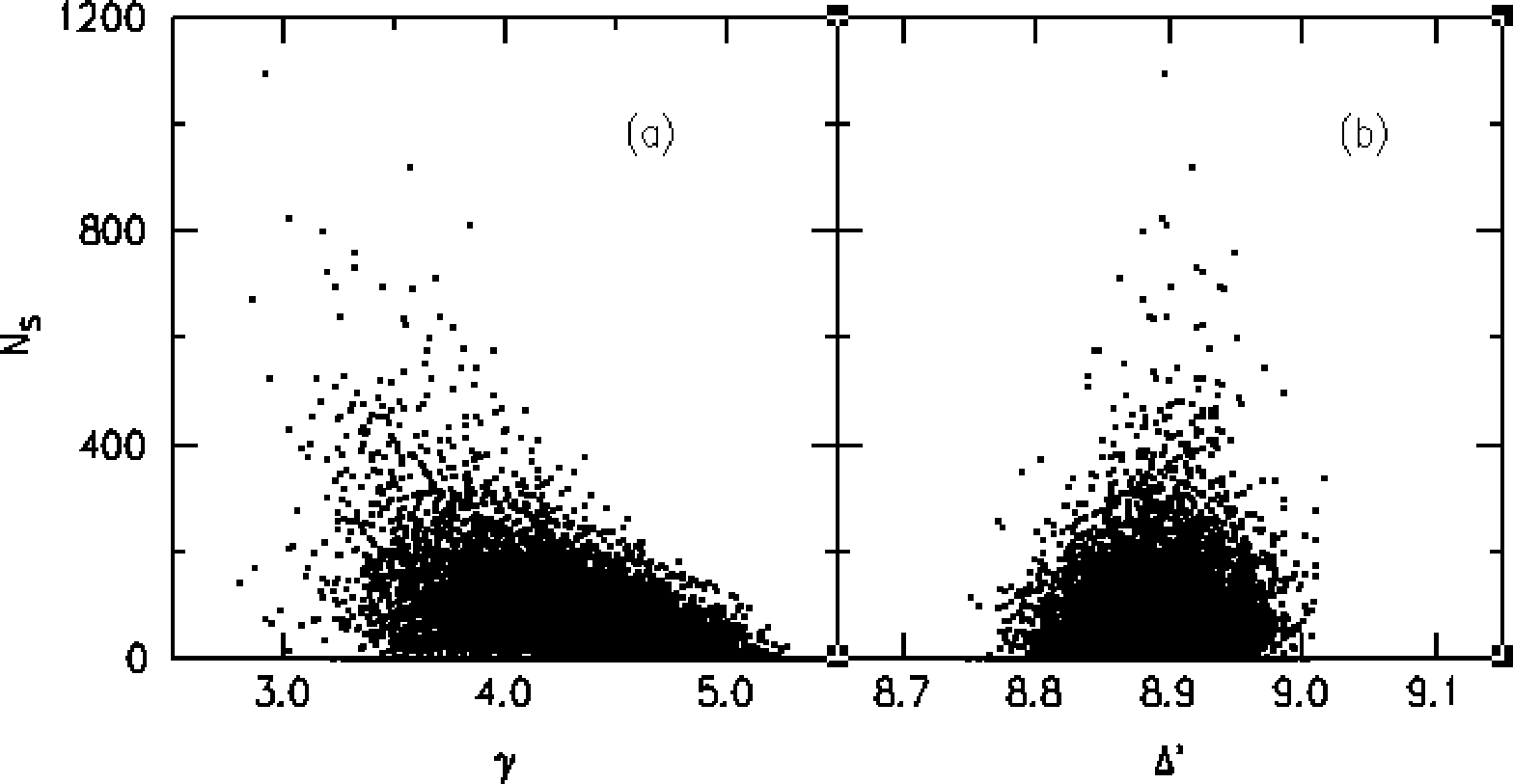,width=5in}}
\caption{$N_S$ vs. $\gamma$ (a) and $\Delta'$ (b) for all the $6\times6$
compact structures.}
\label{zscore}
\end{figure}

Will highly designable structures also fold relatively fast? This question
is addressed in detail in Ref.~\cite{regis} (see also Ref.~\cite{buch2}). A
quantity often used to
measure how much a sequence is ``proteinlike'' is the $Z$ score,
\begin{equation}
Z=\frac{\Delta}{\Gamma},
\label{z}
\end{equation}
where $\Delta$ is the average energy difference between the ground state and
all other states and $\Gamma$ is the standard deviation of the energy
spectrum. $Z$ score was first introduced in the inverse folding problem
\cite{z-score} and later used in protein design \cite{z-score2}. It has been
shown in the context of the Random Energy Model \cite{derrida} that $Z$ score
is related to $T_f/T_g$ where $T_f$ is the folding temperature and $T_g$ the
glass transition temperature \cite{gold}. We have found a good and negative
correlation between the folding time and the $Z$ score of the compact
structure energy spectrum \cite{regis}. In the context of the PH model
(\ref{ham1}), for a sequence $\vec{h}$ and its ground state $\vec{s}$,
\begin{eqnarray}
\label{Del}
\Delta &=& \sum_i h_i (s_i-<s_i>), \\
\Gamma &=& \sqrt{\sum_{ij} h_ih_j c_{i,j}},
\label{Gam}
\end{eqnarray}
where $c_{ij}$ is given by Eq.~(\ref{cij}).
So in principle for every structure $\vec{s}$ one can maximize the $Z$ score
with respect to $\vec{h}$ to get the ``best'' or ``ideal'' sequence for
$\vec{s}$ that gives the highest $Z$ score, $Z_S$. It is however much easier
to obtain a lower bound ${Z_S}'$ for $Z_S$ by letting $\vec{h}=\vec{s}$:
${Z_S}'=\Delta'/\Gamma'$ with
\begin{eqnarray}
\label{Del_p}
\Delta' &=& \sum_i (s_i^2 - s_i<s_i>), \\
\Gamma' &=& \gamma/2,
\label{Gam_p}
\end{eqnarray}
where $\gamma$ is given by Eq.~(\ref{gamma}). In Fig.~\ref{zscore}b, the
$\Delta'$ for all the $6\times6$ compact structures are plotted against
$N_S$ for the PH model. There is little if any correlation between $N_S$
and $\Delta'$ for the $6\times6$ PH model. Thus, correlations between
$N_S$ and $Z'$ in this model come mainly from the one between $N_S$ and
$\Gamma'=\gamma/2$ (Fig.~\ref{zscore}a). So a large $Z'$ is a necessary
but not sufficient condition for a structure to have a large $N_S$.

\section{Summary}
We have demonstrated with simple models that structures are very different
in terms of their designability and that high designability leads to
thermodynamic stability, ``proteinlike'' structural motifs, and foldability.
Highly designable structures emerge because of an {\it asymmetry} between
the sequence and the structure ensembles. Our results are rather robust and
have been demonstrated recently in larger lattice models \cite{433} and in
off-lattice models \cite{miller}.
A broad distribution of designability has also been found in RNA secondary
structures \cite{schu}. However, the set of all sequences designing a
good structure, instead of forming a compact ``Voronoi polytope'' like in
proteins, forms a ``neutral network'' percolating the entire space
\cite{schu}. It would be interesting to study the similarities and
differences of the two systems. Finally, our picture indicates that the
properties of the proteinlike sequences are intimately coupled to that of
the proteinlike (i.e. the highly designable) structures; the picture
unifies various aspects of the two special ensembles. It also suggests
that understanding the emergence and properties of the highly designable
structures is a key to the protein folding problem.

This work was done in collaboration with Hao Li, Ned Wingreen, R\'egis
M\'elin, Robert Helling, and Jonathan Miller. I am grateful to Jeannie
Chen for her critical reading of the manuscript.

\newpage

%%%%%%%%%%%%FIGURE 1%%%%%%%%%%%%%%%

%%%%%%%%%%%%%FIGURE 2%%%%%%%%%%%%%%%

%%%%%%%%%%%%FIGURE 3%%%%%%%%%%%%%%%

%%%%%%%%%%%%%%FIGURE 4%%%%%%%%%%%%%

%%%%%%%%%%%%%%%%FIGURE 5%%%%%%%%%%%%%%%%%

%%%%%%%%%%%%%%%%%%FIGURE 6%%%%%%%%%%%%%%

%%%%%%%%%%%%%%%%%FIGURE 7%%%%%%%%%%%%%%%

%%%%%%%%%%%%%FIGURE 8%%%%%%%%%%%%%%%%

%%%%%%%%%%%%%FIGURE 9%%%%%%%%%%%%%%%%

%%%%%%%%%%%%%FIGURE 10%%%%%%%%%%%%%%%%

%%%%%%%%%%%%%FIGURE 11%%%%%%%%%%%%%%%%

%%%%%%%%%%%%%FIGURE 12%%%%%%%%%%%%%%%%

%%%%%%%%%%%%%FIGURE 13%%%%%%%%%%%%%%%%

%%%%%%%%%%%%%FIGURE 14%%%%%%%%%%%%%%%%


\begin{thebibliography}{9}
\bibitem{foot-des}
This question has been addressed by a number of authors from different
viewpoints. For example, Finkelstein and co-workers took a purely
energetic point of view \cite{fink}, and argued that a structure with lower
energy (averaged over random sequences) will have a larger chance of being
the ground state of more sequences. We will see that this is not the case
in our study where all the structures we consider are maximally compact
and have the same average energy. A more closely related approach is by
Govindarajan and Goldstein who studied the ``foldability'' of
structures \cite{gold0} which is closely related to the designability.
\bibitem{foot-e}
The system is surrounded by water. The energy $e_{\nu\mu}$ is the relative
energy of forming a $\nu$-$\mu$ contact in water. That is that one can
think of $e_{\nu\mu}=E_{\nu\mu}+E_{ww}-E_{\nu w}-E_{\mu w}$, where the $E$'s
are ``absolute'' energies and the subscript $w$ denotes water molecules.
\bibitem{hydroph}
The hydrophobic interaction is of the order $k_B$T (T being the room
temperature). That is that the energy gain for burying a very hydrophobic
amino acid is about a few $k_B$T. See, e.g. Ref.~\cite{protein} for values
of hydrophobicity for the 20 amino acids.
\end{thebibliography}
\end{document}